\documentclass[journal=jpclcd,manuscript=article,layout=twocolumn]{achemso}

\usepackage[version=3]{mhchem} % Formula subscripts using \ce{}
\usepackage{amsmath}
\usepackage{amssymb}
\usepackage{color}
\usepackage{graphicx}
\usepackage{tabularx}
\usepackage{dcolumn} % Align table columns on decimal point
\usepackage{bm}      % bold math
\usepackage{array}   % table making
\usepackage{amsfonts}
\usepackage[bookmarks=false,colorlinks,citecolor=red]{hyperref}
\usepackage{multirow}
\usepackage{caption}
\author{Hongxia Hao}
\affiliation{
Department of Chemistry, Brown University, Providence, RI 02912
}
\author{James Shee}
\affiliation{
Department of Chemistry, Columbia University, New York, NY 10027
}
\author{Shiv Upadhyay}
\affiliation{Department of Chemistry, University of Pittsburgh, Pittsburgh, PA 15260
}
\author{Can Ataca}
\affiliation{
Department of Physics, University of Maryland - Baltimore County, Baltimore, MD 21250
}
\author{Kenneth D. Jordan}
\affiliation{Department of Chemistry, University of Pittsburgh, Pittsburgh, PA 15260
}
\author{Brenda M. Rubenstein}
\email{brenda_rubenstein@brown.edu}
\phone{(603) 661-2160}
\affiliation{
Department of Chemistry, Brown University, Providence, RI 02912
}

\title[]
  {Accurate Predictions of Electron Binding Energies of Dipole-Bound Anions via Quantum Monte Carlo Methods}


\begin{document}
%%%%%%%%%%%%%%%%%%%%%%%%%%%%%%%%%%%%%%%%%%%%%%%%%%%%%%%%%%%%%%%%%%%%%
%% The manuscript does not need to include \maketitle, which is
%% executed automatically.  The document should begin with an
%% abstract, if appropriate.  If one is given and should not be, the
%% contents will be gobbled.
%%%%%%%%%%%%%%%%%%%%%%%%%%%%%%%%%%%%%%%%%%%%%%%%%%%%%%%%%%%%%%%%%%%%%
%%%%%%%%%%%%%%%%%%%%%%%%%%%%%%%%%%%%%%%%%%%%%%%%%%%%%%%%%%%%%%%%%%%%%
%% The "tocentry" environment can be used to create an entry for the
%% graphical table of contents.
%%%%%%%%%%%%%%%%%%%%%%%%%%%%%%%%%%%%%%%%%%%%%%%%%%%%%%%%%%%%%%%%%%%%%
\makeatletter
\setlength\acs@tocentry@height{5cm}
\setlength\acs@tocentry@width{3.8cm}
\makeatother

\begin{tocentry}
\includegraphics[width=5.05cm,height=5.05cm,keepaspectratio]{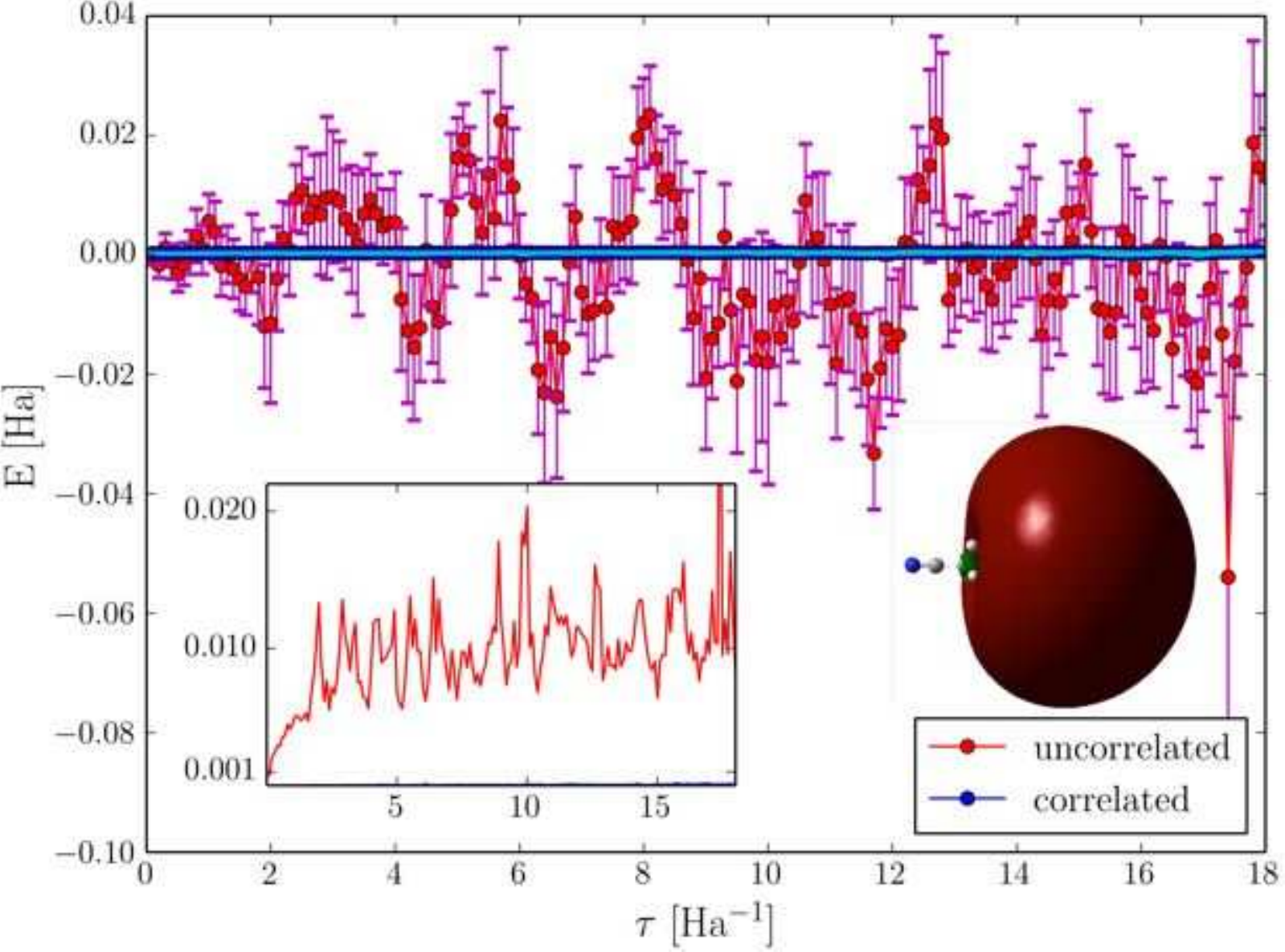}
%{Some journals require a graphical entry for the Table of Contents. This should be laid out ``print ready'' so that the sizing of the text is correct. Inside the \texttt{tocentry} environment, the font used is Helvetica 8\,pt, as required by \emph{Journal of the American Chemical Society}.The surrounding frame is 9\,cm by 3.5\,cm, which is the maximum permitted for  \emph{Journal of the American Chemical Society} graphical table of content entries. The box will not resize if the content is too big: instead it will overflow the edge of the box.This box and the associated title will always be printed on a separate page at the end of the document.}

\end{tocentry}
\begin{abstract}
 Neutral molecules with sufficiently large dipole moments can bind electrons in diffuse nonvalence orbitals with most of their charge density far from the nuclei, forming so-called dipole-bound anions. Because long-range correlation effects play an important role in the binding of an excess electron and overall binding energies are often only of the order of 10-100s of wave numbers, predictively modeling dipole-bound anions remains a challenge. Here, we demonstrate that quantum Monte Carlo methods can accurately characterize molecular dipole-bound anions with near threshold dipole moments. We also show that correlated sampling Auxiliary Field Quantum Monte Carlo is particularly well-suited for 
resolving the fine energy differences between the neutral and anionic species. These results shed light on the fundamental limitations of quantum Monte Carlo methods and pave the way toward using them for the study of weakly-bound species that are too large to model using traditional electron structure methods. 
\end{abstract}
%BR Note: Abstract is at 150 words; have to be vague on the techniques

%%%%%%%%%%%%%%%%%%%%%%%%%%%%%%%%%%%%%%%%%%%%%%%%%%%%%%%%%%%%%%%%%%%%%
%% Start the main part of the manuscript here.
%%%%%%%%%%%%%%%%%%%%%%%%%%%%%%%%%%%%%%%%%%%%%%%%%%%%%%%%%%%%%%%%%%%%%
Dipole-bound anions are intriguing species that bind excess electrons via their molecular dipole moments \cite{Simons_JPCA_2008,Jordan_Wang_Ann_Rev_2003}. As the charge-dipole attraction is governed by a long-range potential that behaves as $1/r^{2}$ at large $r$, dipole-bound electrons are delicately bound in diffuse orbitals with most of their charge density located far from the atomic centers of their parent molecules\cite{Jordan_Luken_JCP_1976,Gutowkshi_Jordan_PRA_1996,Barnett_JCP_1988}. Within the Born-Oppenheimer approximation, the critical dipole moment necessary for binding an electron is 1.625 D, \cite{Fermi_Teller_PhysRev_1947,Turner_Anderson_PhysRev_1968,Crawford_ProcPhysSoc_1967} but increases to 2.5 D or larger when corrections to the Born-Oppenheimer approximation are made\cite{Garrett_CPL_1970,Garrett_PRA_1971,Lykke_PRL_1984,Desfrancois_Schermann_PRL_1994}. Beyond being ``doorways'' to the formation of valence-bound anions\cite{Hendricks_JCP_1998,Compton_JCP_1996,Desfrancois_JCP_1996,Desfrancois_JPCA_1998}, dipole-bound anions may be key contributors to the diffuse interstellar bands, a set of absorption peaks emanating from the interstellar medium whose source has yet to be conclusively identified \cite{Sarre_MNRAS_2000, Sarre_JMS_2006, Guthe_AstroPhysJ_2001, Maier_Nagy_APJ_2011, McCall_AstroPhysJ_2002, Fortenberry_JCP_2011, Larsson_RepProgPhys_2012}. The sheer experimental challenge of resolving the exceedingly small binding energies of such fragile species has motivated spectroscopists to produce dipole-bound species via electron attachment\cite{Hendricks_JCP_1996,Buytendyk_JCP_2016} and Rydberg electron transfer,\cite{Desfrancois_Schermann_PRL_1994,Desfrancois_IntJModPhysB_1996,Hammer_JCP_2003} and to study them via field detachment and photoelectron spectroscopy \cite{Wang_RevSciInstr_1999}. From the theoretical perspective, dipole-bound anions are of particular interest because they pose a formidable challenge for \textit{ab initio} methods -- only high levels of theory, such as coupled cluster theories combined with large, flexible basis sets, are capable of accurately predicting dipole-bound anion electron binding energies that are often of the order of just a few hundred wave numbers in magnitude\cite{Gutowski_PRA_1996,Jordan_Wang_Ann_Rev_2003,Gutowski_Simons_IJQC_1997,Nooijen_JCP_1995,Gutowkshi_Jordan_PRA_1996,Gutzev_CPL_1997}. However, these highly accurate methods scale steeply with system size, severely restricting the size of systems to which they can be applied.  

Herein, we explore the accuracy with which Diffusion Monte Carlo (DMC)\cite{Foulkes_Rajagopal_RMP_2001,Morales_JCTC_2012,Petruzielo_JCP_2012} and Auxiliary Field Quantum Monte Carlo (AFQMC),\cite{Motta_Zhang_2018,Suewattana_Zhang_PRB_2007,Purwanto_JCP_2015,AlSaidi_PRB_2006,AlSaidi_JCP_2006_2,AlSaidi_JCP_2007} two highly accurate, stochastic methods that scale as only $O(N^{3})-O(N^{4})$ with system size, can model dipole-bound anions, with the aim of uncovering a new set of approaches for modeling dipole- and correlation-bound\cite{Voora_JCP_2017} anions of large molecules. Interestingly, despite the different approximations they employ, we find that both methods reproduce experimental results and distinguish molecules that bind an extra electron from those that do not. Furthermore, we find that a newly-developed correlated sampling AFQMC approach (C-AFQMC) \cite{Shee_Reichman_JCTC_2017} is particularly well-suited for the task of studying energy differences involving weakly-bound species and converges electron binding energies orders of magnitude faster than stochastic methods that do not employ such sampling.

To gauge the viability of characterizing dipole-bound anions using QMC methods, we compute vertical electron affinities for several systems known to form dipole-bound anions. Vertical electron affinities may be obtained by taking the difference between the energies of the neutral and anionic species both calculated at the neutral geometry. However, since the dipole-bound excess electron makes almost no impact on the geometry of the molecule, we can equate these electron affinities to the electron binding energies (EBEs), which are typically defined using the geometry of the anion. In the following paragraphs, we summarize the calculations performed; further details may be found in the Supplemental Information. 

In order to compute the EBEs, the neutral geometries were first optimized using the MP2 method\cite{Moller_Plesset_PR_1934} together with the aug-cc-pVDZ basis set\cite{Dunning_JCP_1989,Kendall_JCP_1992} in Gaussian 09.\cite{Frisch_Head-Gordan_CPL_1990,Frisch_Head-Gordan_CPL_1990_2,Head-Gordan_Pople_CPL_1988,Saebo_CPL_1989,Head-Gordan_CPL_1994} Hartree-Fock (HF) wave functions were then generated in Gaussian 09, GAMESS,\cite{Schmidt_JCC,Gordon_2005} or NWChem\cite{Valiev_2010} for use as trial wave functions, which guide sampling and curb the growth of the sign/phase problem, an exponential decay in the signal to noise ratio, in DMC and AFQMC \cite{Anderson_JCP_1975,Ceperley_Fermion_Nodes}. In order to obtain stable dipole-bound anions, it is essential to use flexible basis sets with very diffuse basis functions. Here, we use the aug-cc-pVDZ basis set augmented with a set of diffuse $s$ and $p$ functions, and in one case, also $d$ functions, located near the positive end of the dipole\cite{Gutowkshi_Jordan_PRA_1996}. DMC\cite{Foulkes_Rajagopal_RMP_2001,Needs_JPCM_2010,Toulouse_Review} and AFQMC\cite{Motta_Zhang_2018, Shiwei_Review_2013} are performed on all of the species studied. DMC calculations were conducted using the CASINO package\cite{Needs_JPCM_2010,Drummond_Needs_PRB_2005}. Averages were obtained by sampling the configurations of 4650 walkers for 500,000 or more propagation steps. In order to obtain DMC energies in the zero time-step ($\Delta \tau \rightarrow 0$) limit, DMC simulations were conducted at three different time step sizes and then linearly extrapolated to zero time-step to yield the final values reported. 

Because the calculation of dipole-bound anion vertical binding energies involves energy differences between two species with identical molecular geometries, we employed C-AFQMC \cite{Shee_Reichman_JCTC_2017} for the majority of our AFQMC calculations. In this approach, differences between two quantities normally computed separately using independently generated auxiliary field configurations are instead computed using the same set of configurations. For sufficiently similar systems like many dipole-bound anions and their parent neutral molecules, this can result in a systematic cancellation of errors, which markedly reduces the variance associated with calculated observables. In our calculations, a set of randomly-seeded repeat simulations were initialized, and after an initial equilibration period, the mean and standard error of the cumulative averages were computed among the repeats.  Convergence is attained when the mean is visually observed to plateau, and when the statistical error falls below a target threshold.  The $\Delta\tau \rightarrow 0$ limit was estimated via linear extrapolations using simulations performed at $\Delta \tau=0.01$ and $\Delta \tau=0.005$ a.u. 

%%%% begin table----------------------------------------------
\begin{table*}
%\scriptsize %The standard font size switches are: \tiny, \scriptsize,  \footnotesize, \small, \normalsize, \large, \Large, \LARGE, \huge, and \Huge.
\centering
%\captionsetup{font=normal}
\footnotesize{\small}   
\label{t1}
\resizebox{\textwidth}{!}{%
\begin{tabular}{|c|c|c|c|c|c|c|}
\hline 
&\multirow{2}*{} & \multicolumn{5}{c|}{{\bf Electron Binding Energy (cm$^{-1}$)} }\\
\cline {3-7}
& {\bf Dipole Moment (D)} & {\bf Experiment} & {\bf $\Delta$SCF\textsuperscript{\emph{a}}} & {\bf CCSD(T)\textsuperscript{\emph{a}}} & {\bf DMC\textsuperscript{\emph{b}}} & {\bf C-AFQMC\textsuperscript{\emph{c}}}\\ 
\hline
{\bf SO} &1.55\cite{database_dipole_moment} &NOT-BOUND &-3.84  &-4.13 &-308.20 $\pm$ 70.82 &-4.54 $\pm$ 0.64\\
\hline
{\bf HCN} &2.98\cite{database_dipole_moment} &13\cite{Ard_Stepanian_CPL_2009} &11.00 &7.44 &46.17 $\pm$ 45.30 &10.80 $\pm$ 2.95\\
\hline
{\bf CH$_2$CHCN}  &3.87\cite{database_dipole_moment} &56 - 87\cite{Desfrancois_Schermann_PRL_1994,Desfrancois_PRA_1995} &43.30 &61.87 &106.63 $\pm$ 58.12 &65.70 $\pm$ 11.03 \\
\hline
{\bf CH$_3$CN} &3.92\cite{database_dipole_moment} &93 - 145\cite{Desfrancois_Schermann_PRL_1994,Desfrancois_PRA_1995} &50.83 &103.00 &93.83 $\pm$ 36.21 &95.85 $\pm$ 9.73 \\
\hline
{\bf C$_3$H$_2$} &4.14\cite{Gottlieb_Oswald_JCP_1993} &170 $\pm$ 50\cite{Yokoyama_Gutowski_JCP_1996} &54.61 &162.08 &151.22 $\pm$ 64.25\textsuperscript{\emph{d}} &132.45 $\pm$ 9.43\textsuperscript{\emph{e}}\\
\hline
{\bf C$_3$H$_2$O$_3$} &4.55\cite{database_dipole_moment} &194 $\pm$ 24\cite{Hammer_Jordan_JCP_2004} &103.13 &163.31 &213.98 $\pm$ 116.15 &157.70 $\pm$ 17.96 \\
\hline
\end{tabular}}
\caption{EBEs and dipole moments of selected species from experiment and Self-Consistent Field [HF], Coupled Cluster [CCSD(T)], DMC, and C-AFQMC calculations.}
\textsuperscript{\emph{a}} HF and CCSD(T) calculations were preformed using Gaussian 09.
\textsuperscript{\emph{b}} DMC calculations of the anion are based on unrestricted Hartree Fock (UHF) trial wave functions obtained from Gaussian 09 or GAMESS.
\textsuperscript{\emph{c}} AFQMC calculations were based on UHF trial wave functions obtained from NWChem.
\textsuperscript{\emph{d}} The DMC calculations on the C$_{3}$H$_{2}$ dipole-bound anion used a restricted open-shell Hartree Fock (ROHF) trial wave function.
\textsuperscript{\emph{e}} To be consistent, the C-AFQMC calculations on the C$_{3}$H$_{2}$ dipole-bound anion were also based on an ROHF trial wave function. 
\label{t1}
\end{table*}
%%%% end table----------------------------------------------

While DMC has been successfully employed to calculate cluster binding energies,\cite{Xu_JPCA_2010,Wang_JPCA_2013} here we analyze its performance predicting the electron binding energies of molecular dipole-bound anions. As an initial test, we examined whether DMC could faithfully predict whether a given species binds an extra electron or not. Thus, in addition to considering several species known to form dipole-bound anions, we also consider SO, a molecule with a dipole moment of 1.55 D, which is below the threshold required for binding. Our results are summarized in Table \ref{t1}. DMC calculations correctly predicted that all of the species studied, except SO, would form dipole-bound anions. 

\begin{center}
\begin{figure}[htbp]
\includegraphics[width=8.5cm]{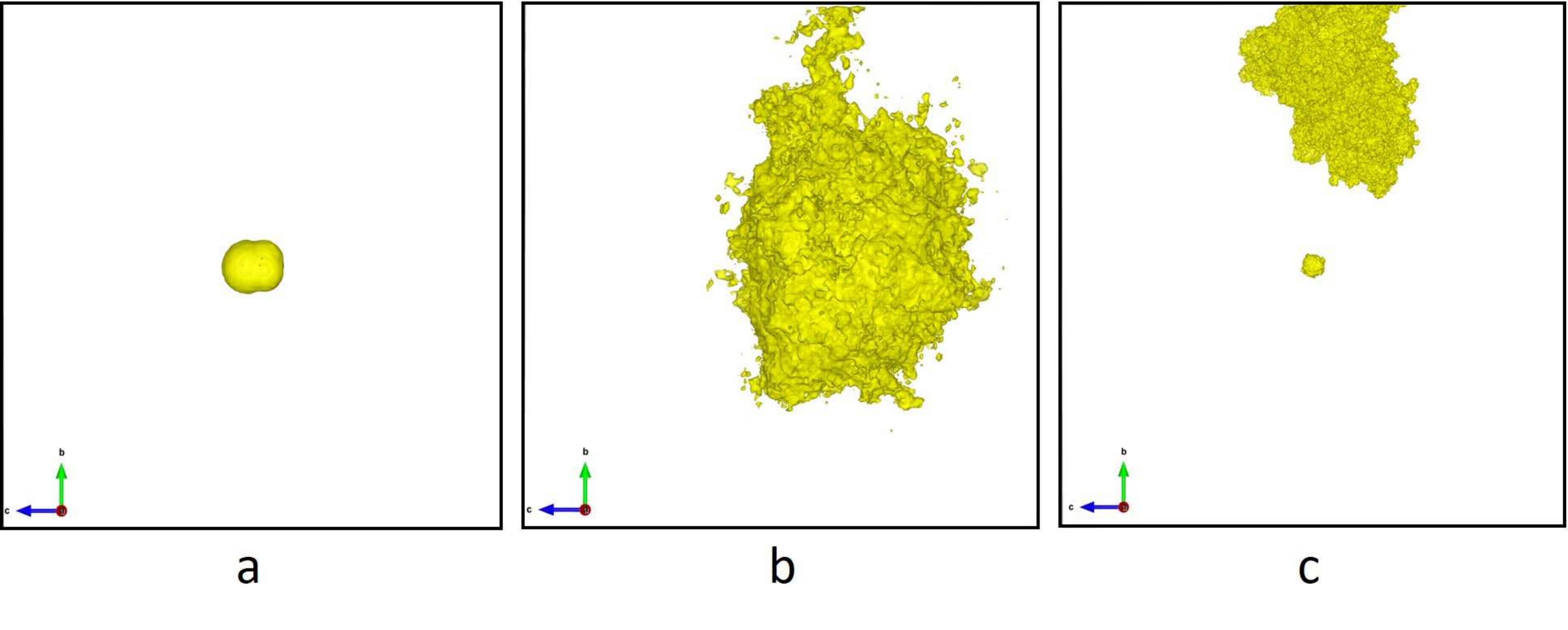}
\captionsetup{font=footnotesize, justification=raggedright, singlelinecheck=false}
\caption{DMC charge densities of (a) neutral CH$_3$CN, (b) the CH$_3$CN anion, and (c) SO plus an extra electron. The isosurface values taken for each of these plots are $4 \times 10^{-14}$ e/$\mbox{\AA}^3$, $4\times 10^{-14}$ e/$\mbox{\AA}^3$, and $1 \times 10^{-20}$ e/$\mbox{\AA}^3$, respectively. Molecules are placed in the center of the simulation box.}
\label{f1}
\end{figure}
\end{center}

In Figure \ref{f1}, the charge densities of neutral CH$_{3}$CN, the CH$_{3}$CN anion, and the SO molecule plus an extra electron are compared. While the DMC charge density for the neutral CH$_{3}$CN molecule is highly localized in the molecular region, that of the anion is far more diffuse and protrudes continuously out from the positive side of the dipole. The charge density plots of the other dipole-bound anions considered in this work manifest similar features and are reported in the Supplemental Information. In stark contrast, the charge density of SO plus an extra electron consists of two disjoint contributions -- one associated with the neutral molecule and a second representative of an additional unbound electron positioned more than 50 $\mbox{\AA}$ from the molecule. 
%%%% begin figure-----------------------------------------
%%%% end figure------------------------------------------

\begin{center}
\begin{figure}[htbp]
\includegraphics[width=8.5cm]{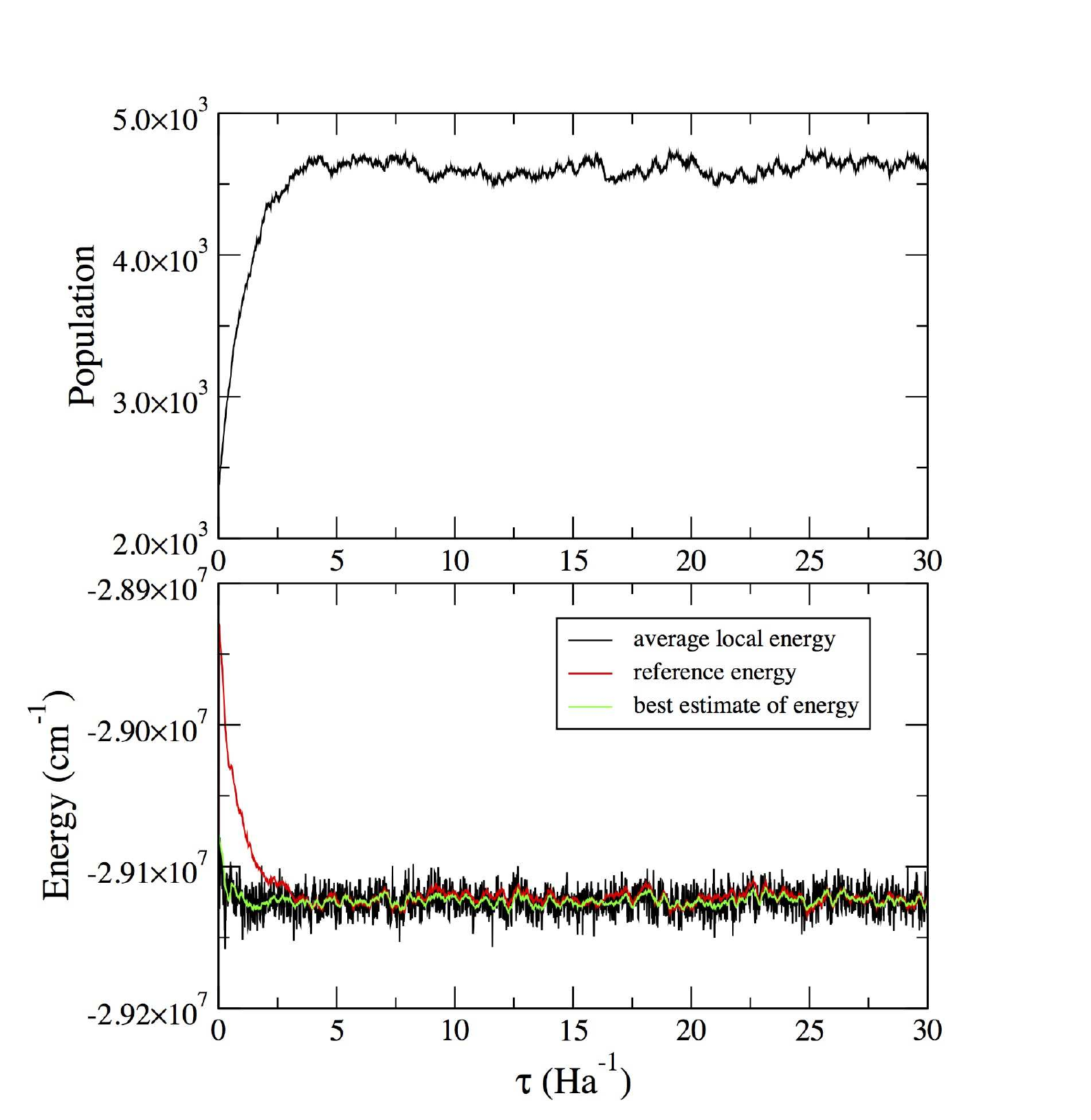}
\captionsetup{font=footnotesize, justification=raggedright, singlelinecheck=false}
\caption{The time evolution of the DMC energy and walker population for the CH$_{3}$CN anion using $\Delta \tau = 0.01$ a.u. with 4650 walkers. Walkers were initialized with a UHF trial wave function expanded in terms of the aug-cc-pVDZ basis with a 7$s$7$p$ set of diffuse gaussian-type orbitals. In the Figure, reference energy refers to $E_{T}$ (see Supplemental Information), the average local energy refers to the local energy averaged over walkers at a given imaginary time, and best estimate of the energy refers to the energy averaged over all samples taken up to a certain imaginary time.}
\label{f2}
\end{figure}
\end{center}

Interestingly, even though the fixed-node error in the energies of the neutral and anion are greater in magnitude than the electron binding energy, DMC calculations using  single Slater determinant trial wave functions provide semi-qualitatively accurate EBEs of dipole-bound anions. Nevertheless, obtaining quantitatively accurate EBEs with the DMC approach employed here would be too computationally demanding to be practical. As presented in Table \ref{t1}, achieving DMC statistical error bars smaller than the binding energies can require hundreds of thousands to millions of DMC iterations, even starting from a well-optimized variational wave function. For example, as depicted in Figure \ref{f2}, statistical fluctuations in the energy of the CH$_{3}$CN anion simulated with 4650 walkers hover around 65000 cm$^{-1}$, meaning that over 500,000 samples must be taken to achieve on the order of 100 cm$^{-1}$ error bars. Thus, while DMC EBEs that agree with experimental measurements and coupled cluster calculations can be gleaned from the noise, it is only with error bars that are still too large to make definitive statements and at great computational expense.
%%%% begin figure----------------------------------------------
%%%% end figure---------------------------------------------- 
\begin{center}
\begin{figure*}[htbp]
\includegraphics[width=15cm]{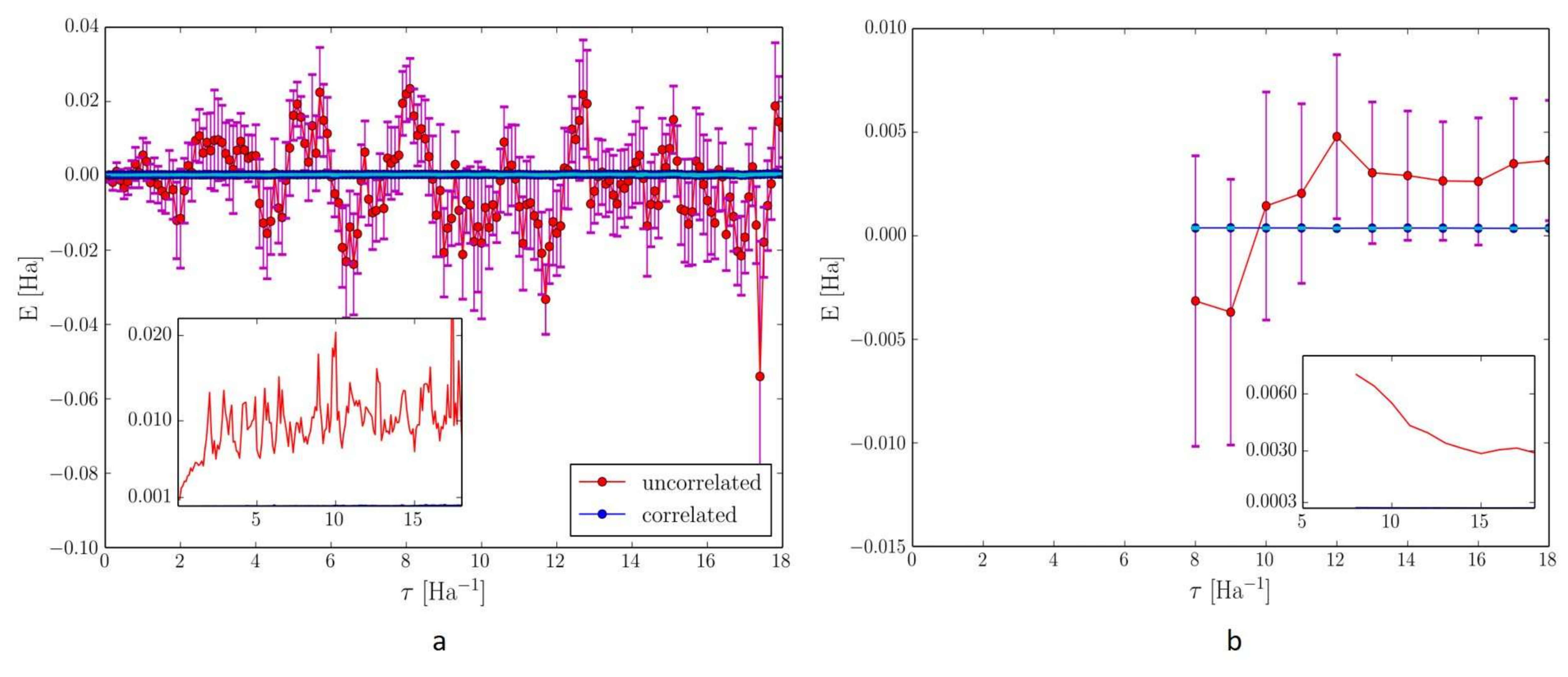}
\captionsetup{font=footnotesize, justification=raggedright, singlelinecheck=false}
\caption{Comparison of energies from AFQMC calculations with and without correlated sampling for CH$_{3}$CN using the aug-cc-pvdz+7$s$7$p$ basis, an HF reference wave function, $\Delta \tau=0.01$ a.u., 384 walkers per simulation, and 17 repeated simulations. (a) Mean values of the EBE (circles) among the repeats at each $\tau$ along the imaginary-time propagation; (b) Mean values of the cumulative averages taken for $\tau > 8$ a.u. The error bars give the standard errors, defined as the standard deviation times $\frac{1}{\sqrt[]{N_r}}$, where $N_r$ is the number of simulation repeats, and are plotted in the insets for clarity.}
\label{f3}
\end{figure*}
\end{center}
%%%% end figure----------------------------------------------
One stochastic technique capable of scaling to dipole-bound anions of large molecules with a substantial reduction in statistical noise is C-AFQMC. As shown in Table \ref{t1}, C-AFQMC error bars are at least an order of magnitude smaller than DMC error bars using
a similar number of samples, but using two orders of magnitude shorter projection time (2000 a.u. was used to obtain most DMC results; 20 a.u. was used to obtain the C-AFQMC results). The C-AFQMC results presented were moreover obtained starting from just HF wave functions. One might question whether the source of this improvement stems from the AFQMC algorithm or from the use of correlated sampling. Figure \ref{f3}, which depicts the energy as a function of imaginary projection time for the CH$_{3}$CN molecule, demonstrates that the improvement may be attributed to \textit{both} sources. Uncorrelated AFQMC simulations are accompanied by statistical fluctuations on the order of $10^{4}$ cm$^{-1}$, which are smaller than the $10^{5}$ cm$^{-1}$ fluctuations associated with the EBE values from the DMC simulations (see Figure \ref{f2}). C-AFQMC calculations are accompanied by almost imperceptible statistical fluctuations on the order of 22 cm$^{-1}$. Thus, AFQMC's sampling innovations yield meaningful gains above DMC, yet it is the use of correlated sampling that yields, \textit{by far}, the largest improvements. 

As shown in Table \ref{t1}, C-AFQMC yields energies with sufficiently small error bars that meaningful comparisons can now be made against coupled cluster calculations and experiment. In general, the EBEs predicted by C-AFQMC are within error bars of both experimental and previous coupled cluster single, doubles, and perturbative triples (CCSD(T)) calculations. This demonstrates that phaseless approximation errors are mild and that for dipole-bound anions AFQMC is as accurate as CCSD(T). The C$_{3}$H$_{2}$ anion, however, stands as one cautionary tale. For the C$_{3}$H$_{2}$ anion, we located two different UHF solutions, neither of which proved suitable for trial functions in QMC calculations (see the
Supplemental Information for more details). For this reason, we employed an ROHF trial wave function for the C$_{3}$H$_{2}$ anion instead. This example suggests that care must be taken when selecting trial wave functions for and ultimately simulating larger molecules that may have many competing low-lying states.

In this work, we have demonstrated that low-scaling QMC methods, and in particular, C-AFQMC, are capable of resolving the fine energy differences required to accurately predict electron binding energies of dipole-bound anions. Electron binding energies within wave numbers of previous coupled cluster and experimental results were obtained for HCN, CH$_{2}$CHCN, CH$_{3}$CN, C$_{3}$H$_{2}$, and C$_{3}$H$_{2}$O$_{3}$. Our results demonstrate that, while uncorrelated DMC and AFQMC methods can \textit{qualitatively} describe dipole-bound species, only correlated sampling techniques such as C-AFQMC are capable of achieving \textit{quantitative} accuracy within  computationally tractable amounts of time. The success of C-AFQMC in this work beckons for the further development of correlated DMC methods capable of resolving the molecular energy differences that lie at the heart of all chemical processes. These findings pave the way toward using stochastic methods to study the much larger polycyclic aromatic hydrocarbons (PAH) \cite{Voora_JPCL_2015} and long-chain carbon anions\cite{Tulej_AJL_1998} thought to contribute to the diffuse interstellar bands, as well as correlation-bound anions\cite{Voora_Jordan_JPCL_2013,Jordan_Sommerfeld_JCP_2017} and weakly bound clusters \cite{Zen_PNAS_2017} whose size puts them beyond reach of most high accuracy methods.

%%%%%%%%%%%%%%%%%%%%%%%%%%%%%%%%%%%%%%%%%%%%%%%%%%%%%%%%%%%%%%%%%%%%%
%% The "Acknowledgement" section can be given in all manuscript
%% classes.  This should be given within the "acknowledgement"
%% environment, which will make the correct section or running title.
%%%%%%%%%%%%%%%%%%%%%%%%%%%%%%%%%%%%%%%%%%%%%%%%%%%%%%%%%%%%%%%%%%%%%
\begin{acknowledgement}

H.H., J.S., and B.R. acknowledge Lai-Sheng Wang, Yuan Liu, G. Stephen Kocheril, Shiwei Zhang, and David Reichman for their ongoing support and insights. H.H. and B.R. acknowledge support from the U.S. Department of Energy by Lawrence Livermore National Laboratory under Contract DE-AC52-07NA27344, 15-ERD-013 and NSF grant DMR-1726213. S.U. and K.D.J. acknowledge the support of NSF grant CHE-1762337. This research was conducted using computational resources and services at the Center for Computation and Visualization, Brown University, the University of Pittsburgh's Center for Research Computing, and the Extreme Science and Engineering Discovery Environment (XSEDE). 

\end{acknowledgement}

%%%%%%%%%%%%%%%%%%%%%%%%%%%%%%%%%%%%%%%%%%%%%%%%%%%%%%%%%%%%%%%%%%%%%
%% The same is true for Supporting Information, which should use the
%% suppinfo environment.
%%%%%%%%%%%%%%%%%%%%%%%%%%%%%%%%%%%%%%%%%%%%%%%%%%%%%%%%%%%%%%%%%%%%%
%\begin{suppinfo}

%This will usually read something like: ``Experimental procedures and
%characterization data for all new compounds. The class will
%automatically add a sentence pointing to the information on-line:

%\end{suppinfo}

%%%%%%%%%%%%%%%%%%%%%%%%%%%%%%%%%%%%%%%%%%%%%%%%%%%%%%%%%%%%%%%%%%%%%
%% The appropriate \bibliography command should be placed here.
%% Notice that the class file automatically sets \bibliographystyle
%% and also names the section correctly.
%%%%%%%%%%%%%%%%%%%%%%%%%%%%%%%%%%%%%%%%%%%%%%%%%%%%%%%%%%%%%%%%%%%%%
\bibliography{references}

\end{document}